\documentstyle[12pt,epsfig]{report}
\begin{document}
\newcommand{\beq}{\begin{equation}}
\newcommand{\eeq}{\end{equation}}
\newcommand{\beqa}{\begin{eqnarray}}
\newcommand{\eeqa}{\end{eqnarray}}
\newcommand{\no}{\nonumber}
\newcommand{\lets}{\stackrel{<}{_\sim}}
\newcommand{\ul}{\underline}
\newcommand{\ol}{\overline}
\newcommand{\ra}{\rightarrow}
\newcommand{\ve}{\varepsilon}
\newcommand{\vp}{\varphi}
\newcommand{\vt}{\vartheta}
\newcommand{\dg}{\dagger}
\newcommand{\wt}{\widetilde}
\newcommand{\wh}{\widehat}
\newcommand{\dis}{\displaystyle}
\newcommand{\dfrac}{\displaystyle \frac}
\newcommand{\fsl}{\not\!}
\newcommand{\ben}{\begin{enumerate}}
\newcommand{\een}{\end{enumerate}}
\newcommand{\bfl}{\begin{flushleft}}
\newcommand{\efl}{\end{flushleft}}
\newcommand{\ba}{\begin{array}}
\newcommand{\ea}{\end{array}}
\newcommand{\btab}{\begin{tabular}}
\newcommand{\etab}{\end{tabular}}
\newcommand{\bit}{\begin{itemize}}
\newcommand{\eit}{\end{itemize}}
\newcommand{\be}{\begin{equation}}
\newcommand{\ee}{\end{equation}}
\newcommand{\bearr}{\begin{eqnarray}}
\newcommand{\eearr}{\end{eqnarray}}
\newcommand{\bdm}{\begin{displaymath}}
\newcommand{\edm}{\end{displaymath}}
\newcommand{\bea}{\begin{eqnarray}}
\newcommand{\eea}{\end{eqnarray}}
\newcommand{\fs}{\; \; .}
\newcommand{\co}{\; \; ,}
\newcommand{\nl}{\nonumber \\}
\newcommand{\mscr}[1]{{\mbox{\scriptsize #1}}}
\newcommand{\mtiny}[1]{{\mbox{\tiny #1}}}

\parskip=4pt plus 1pt
\chapter*{}

\begin{center}
{\Large \bf CHIRAL PERTURBATION THEORY
\footnote{To be published in the second edition of the
DA$\Phi$NE Physics Handbook, eds. L. Maiani, G. Pancheri and N. Paver} } \\
\label{Intro}
\thispagestyle{empty}

\vspace*{2cm}
{\bf{J. Bijnens$\,^1$, G. Ecker$\,^2$ and J. Gasser$\,^3$}}

\vspace*{2cm}
${}^{1)}$ NORDITA, Blegdamsvej 17, DK--2100 Copenhagen

${}^{2)}$ Inst. Theor. Physik, Univ. Wien, Boltzmanngasse 5, A--1090 Wien

${}^{3)}$ Inst. Theor. Physik, Univ. Bern, Sidlerstrasse 5, CH--3012 Bern

\end{center}

\newpage

\renewcommand{\thesection}{\arabic{section}}
\renewcommand{\thesubsection}{\arabic{section}.\arabic{subsection}}
\renewcommand{\theequation}{\arabic{section}.\arabic{equation}}
\renewcommand{\thetable}{\arabic{table}}
\renewcommand{\thefigure}{\arabic{figure}}

%before each section
\setcounter{equation}{0}
\setcounter{subsection}{0}

\section{Prelude}

Chiral perturbation theory (CHPT)
is a systematic method to analyse the low--energy structure of the Standard
Model. In particular,  it allows one
 to determine the low--energy behaviour of the Green
functions
built from quark currents, like the electromagnetic form factor of the
mesons or
 the meson--meson scattering amplitudes.
 Matrix elements for
semileptonic decays are calculable from quark current correlators as
well,
because the momenta of the external particles are small with respect to the $W$
 mass, and the weak interactions reduce to the current$\times$current
form in this case. For illustration, we note the following
correspondence:
 \bdm
\begin{tabular}{ccl}
   $\langle0|TA_\mu^i(x)V_\rho^{em}(y)A_\nu^j(z) |0\rangle$&:&electromagnetic
form
factor\\
&&\\
$\langle0|TA_\mu^i(x)A_\nu^k(y)A_\rho^l(z)A_\sigma^m(w)|0\rangle$&:&$\pi\pi$
scattering amplitude \\
&&\\
  $\langle0|TA_\mu^l(x)\bar{u}(y)\gamma_\nu s(y)A_\rho^m(z)|0\rangle$&:&
 $K\rightarrow \pi e \nu \;  \; \mbox{decay}\fs$\\
 \end{tabular} \edm
Here   $A_\mu^i,V_\rho^{em}$ and $Q$ denote the axial and
the electromagnetic current and the charge matrix, respectively,
 \bea
A_\mu^i&=&\bar{q}\gamma_\mu\gamma_5\frac{\lambda^i}{2}q\co
V_\mu^{em}=\bar{q}\gamma_\mu Qq \co\\
Q&=&\frac{1}{3}\mbox{diag}(2,-1,-1)\co \no
\eea
where $q$ collects the $u,d$ and $s$ quark fields,
\beq
q=\left( \begin{array}{c}
                u\\d\\s
          \end{array}\right)\fs
\eeq

It is very convenient to collect all
Green functions in the generating functional whose properties may then be
studied in a
general manner. The construction goes as follows. One equips the
QCD Lagrangian with external $c$--number fields $v_\mu,a_\mu,s$ and $p$,
\beq
{\cal L} = {\cal L}^0_{QCD} + \bar q \gamma^\mu (v_\mu + \gamma_5 a_\mu)q
- \bar q (s - i \gamma_5 p)q ~. \label{eq:QCD}
\eeq
${\cal L}^0_{QCD}$ is the Lagrangian with the masses of the
three light quarks set to zero. The external fields $v_\mu$, $a_\mu$,
$s$ and $p$ are hermitian $3 \times 3$ matrices in flavour space, and
the quark mass matrix
\be
{\cal M} = \mbox{diag}(m_u,m_d,m_s)
\ee
is contained in the scalar field $s$.
The generating functional is the logarithm of the vacuum transition amplitude,
\be\label{gener}
e^{iZ[v,a,s,p]}=\langle0\;\;\mbox{out}|0\;\;\mbox{in}\rangle\co
\ee
and contains the external fields as arguments. Expanding $Z$
in powers of $v_\mu,a_\mu,s$ and $p$  generates the
connected Green functions
of the quark currents. For example, the matrix element for $\pi\pi$
scattering is contained in the quartic term
\bea\label{zpipi}
Z&=&\ldots + \frac{i^3}{4!}\int
dx_1dx_2dx_3dx_4a_\mu^i(x_1)a_\nu^k(x_2)a_\rho^l(x_3)a_\sigma^m(x_4)\times\nl
&&\times\langle0|TA_i^\mu(x_1)A_k^{\nu}(x_2)
A_l^\rho(x_3)A_m^\sigma(x_4)|0\rangle
+\ldots \co
\eea
with $a_\mu=a_\mu^b\frac{\lambda^b}{2}$.

Notice that the Lagrangian (\ref{eq:QCD}) is symmetric under local
$SU(3)_L \times SU(3)_R$ transformations,
\beqa
q & \ra & g_R \, \dfrac{1}{2} (1 + \gamma_5)q +
                  g_L \, \dfrac{1}{2} (1 - \gamma_5)q \co\\
g_{R,L} & \in & SU(3)_{R,L}\no\co
\eeqa
provided the external fields are transformed accordingly,
\beqa\label{localf}
r_\mu=v_\mu+a_\mu  & \ra & g_R r_\mu g^\dagger_R +
                             i g_R \partial_\mu g^\dagger_R \co\nl
l_\mu=v_\mu-a_\mu  & \ra & g_L l_\mu g^\dagger_L +
                             i g_L \partial_\mu g^\dagger_L \co\nl
s + i p & \rightarrow & g_R (s + i p) g^\dagger_L \fs
\eeqa
Here we consider only the case where the fields $a_\mu,v_\mu$ are traceless,
\beq
\langle v_\mu\rangle=\langle a_\mu \rangle = 0\co
\eeq
where $\langle A \rangle$ denotes the trace of the matrix $A$.

In the following we show how, in the framework of CHPT,
the generating functional can be evaluated at low energies in a systematic
manner.

%before each section
\setcounter{equation}{0}
\setcounter{subsection}{0}

\section{Effective low--energy theory}

The generating functional admits an expansion in powers of
external momenta and of quark masses \cite{Wein79,GL1,HL93}. This expansion
is
 most conveniently  carried out in an effective theory, where the quark and
gluon fields of QCD are replaced by a set of pseudoscalar fields which
describe the degrees of freedom of the Goldstone bosons $\pi, K$ and
$\eta$. On the level of this effective theory, the expansion amounts to
a derivative expansion of the effective Lagrangian. Counting the quark
mass as two powers of the momenta, the expansion starts at order
$O(p^2)$ and is of the form
\beq
{\cal L}_{eff}={\cal L}_2+{\cal L}_4+\ldots\;\;\fs
\eeq
The effective Lagrangian generates the corresponding expansion of the
generating functional,
\beq
Z=Z_2+Z_4+\ldots \;\; .
\eeq
This procedure to analyse the low--energy structure of the Green functions is
called ``chiral perturbation theory". If the effective Lagrangian is
chosen properly, the resulting Green functions are identical to the ones in
the Standard Model \cite{Wein79,HL93}.

To see how this works in detail, it is convenient to collect the
pseudoscalar fields in a unitary $3\times 3$ matrix,
\beq
U = \exp{(i\sqrt{2}\Phi/F)}~,\qquad
\Phi = \left( \ba{ccc}
\dfrac{\pi^0}{\sqrt{2}} + \dfrac{\eta_8}{\sqrt{6}} & \pi^+ & K^+ \\*
\pi^- & -\dfrac{\pi^0}{\sqrt{2}} + \dfrac{\eta_8}{\sqrt{6}} &  K^0 \\*
K^- & \ol{K^0} & - \dfrac{2 \eta_8}{\sqrt{6}} \ea \right)~.\no\label{eq:phi}
\eeq
The {\it leading--order term} in the low--energy expansion is then generated by
the
tree diagrams of the non--linear $\sigma$ model coupled to the external fields
$v,a,s$ and $p$,
\be
{\cal L}_2 = \frac{F^2}{4} \langle D_\mu U D^\mu U^\dagger +
             \chi U^\dagger + \chi^\dagger U \rangle \co\label{eq:L2}
\ee
where
\beq
D_\mu U = \partial_\mu U - ir_\mu U + iU l_\mu~, \qquad
\chi = 2 B(s + ip)~. \no\eeq
The effective Lagrangian contains only two parameters at this order in
the energy
expansion:
 the pion decay constant
in the chiral limit,
$F\simeq   93$ MeV,
and  $B$, which is related to the quark condensate,
$\langle 0|\bar u u |0\rangle = - F^2 B[1 + O(m_{quark})]$.
The constant $B$ always appears multiplied by quark masses. At $O(p^2)$, the
product $B m_{quark}$ can be expressed in terms of meson masses, e.g.,
\bea\label{eq:mass}
M^2_{\pi^+} = B (m_u + m_d) + O(m_{quark}^2)\co\\
M^2_{K^+} = B (m_u + m_s) + O(m_{quark}^2)\fs\no
\eea
The fields $U,v,a,s$ and $p$ that occur in the effective Lagrangian are
subject to the following chiral counting rules \footnote{For a
different counting see Ref.~\cite{GCHPT}.},
 \beqa\label{counting}
U & \hspace{.5cm}\sim\hspace{.5cm} & O(p^0) \no\co \\*
D_\mu U, v_\mu, a_\mu &\sim & O(p) \no\co \\*
s,p &\sim & O(p^2) ~.\label{cc}
\eeqa
Therefore, the {\it leading--order} effective Lagrangian ${\cal L}_2$ is of
order
$p^2$. Furthermore, we note that ${\cal L}_2$ is invariant under the local
transformation (\ref{localf}), provided the field $U$ is transformed as
\be\label{localu}
U\rightarrow g_R U g_L^\dagger \fs
\ee

The tree graphs generated by the Lagrangian (\ref{eq:L2}) reproduce the
soft pion theorems for mesons \cite{adler} in a concise manner.
To illustrate, we consider the matrix element for elastic scattering of two
charged pions,
\bea
&&\langle\pi^+(p_2')\pi^-(p_1') \;\;\mbox{out}| \pi^+(p_2)\pi^-(p_1)
 \;\;\mbox{in}\rangle=\nl
&&4(2\pi)^6p_2^0p_1^0\delta^3({\bf p_2' - p_2})\delta^3({\bf p_1'- p_1})
+i(2\pi)^4\delta^4(p_1'+p_2' -p_1-p_2)T(s,t)\co
\eea
where $s=(p_1+p_2)^2,t=(p_1'-p_1)^2$ are the standard Mandelstam variables.
There is only one diagram to  evaluate, see Fig.~1. The result is
\be\label{apipi}
T(s,t)=\frac{s+t-2 M^2_\pi}{F^2}+O(p^4)\fs
\ee
Here the symbol $O(p^4)$ collects terms of order O($p^4,p^2 m_{quark}$,
$m_{quark}^2$).
The result  (\ref{apipi})  is due to Weinberg \cite{Weinpp}
who used current algebra and
PCAC to analyse the Ward identities for the four--point function of the axial
current. The effective Lagrangian calculation is very simple -- the
result is the same.
%\begin{figure}[t]
\begin{figure}
\centerline{\epsfig{file=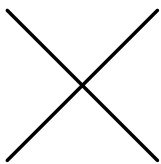,height=2cm}}
%\vspace{4cm}
\caption{The Feynman diagram that generates the leading--order
contribution (\protect\ref{apipi}) to the $\pi\pi$ scattering amplitude.}
\end{figure}

At {\it next--to--leading order}, the generating functional consists of
three terms \cite{GL1} :
\begin{enumerate}
\item[i)] The  one--loop graphs generated by the lowest--order Lagrangian
${\cal L}_2$.
\item[ii )] An explicit local action of order $p^4$.
\item[iii)] A contribution to account for the chiral anomaly.
\end{enumerate}
\noindent
We briefly discuss these three contributions in turn and start with the
one--loop graphs.

 \subsection{One--loop graphs}
For illustration, we consider the electromagnetic form factor $F_V^\pi$ of
the pion,
\beq
\langle\pi^+(p')|V_\mu^{em}(0)|\pi^+(p)\rangle=(p'+p)_\mu F_V^\pi (t)\;\;;
\;\;t=(p'-p)^2\fs
\eeq
We display in Fig.~2 the relevant tree and one--loop graphs
generated by ${\cal L}_2$.

%\begin{figure}[t]
\begin{figure}
\centerline{\epsfig{file=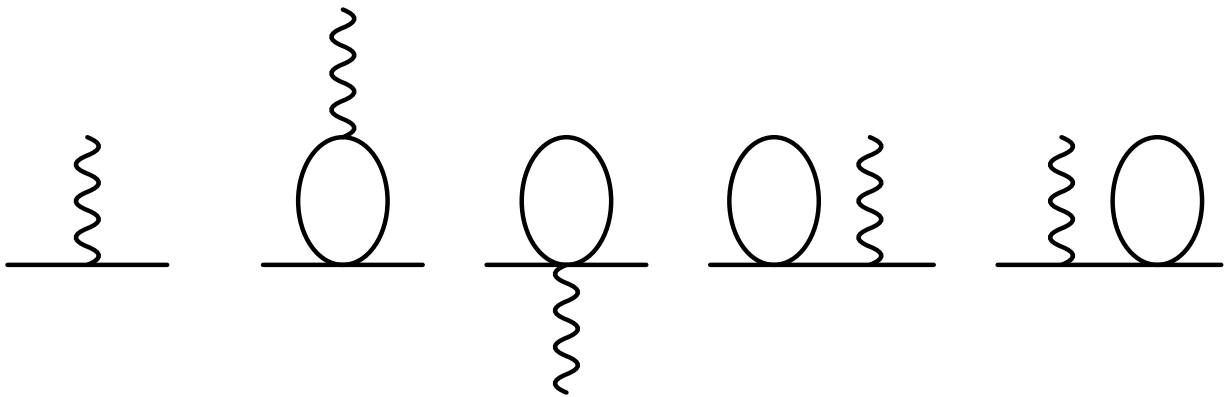,height=5.5cm}}
%\vspace{6cm}
\caption{The tree and one--loop graphs, generated by ${\cal L}_2$,
that contribute to the electromagnetic form factor of the charged pion.
$\pi, K$ and $\eta$ are running in the loops.
The wiggly line denotes the electromagnetic current.}
\end{figure}
The leading--order term (tree graph of ${\cal L}_2$) is trivial,
because $eF_V^\pi(0)$ is the charge of the pion,
\beq
F^{\pi,tree}_V=1\fs
\eeq
 The one--loop graphs  are
ultraviolet divergent, and a regularization scheme is needed. Although the
final result is independent of the regularization chosen, it is very
convenient to use dimensional regularization for the following reason. In this
scheme, the chiral symmetry of the effective theory is maintained, and only
chiral symmetric counterterms are needed to cancel the divergences.
Furthermore, because no new scale is introduced, the one--loop graphs are
suppressed
by two powers of the momenta relative to the tree diagrams \cite{Wein79},
see also subsection \ref{sec:power}.
 The result for the electromagnetic form factor is
 \bea \label{eq100}
 F^{\pi,one\;  loop}_V(t)&=&2\phi(t,M_\pi;d)+\phi(t,M_K;d) \co\nl
\phi(t,M;d)&=&-\frac{tM^{d-4}}{(4\pi)^{d/2}}\frac{\Gamma(2-\frac{d}{2})}{2F^2}
\int_0^1\!dx \;x(1-2x)\left[1-\frac{t}{M^2}x(1-x)\right]^{\frac{d-4}{2}}\co\nl
 \eea
 where we have indicated
the dependence of the loop functions $\phi$ on the dimension
$d$ of space--time.  From this representation of $\phi$ it is explicitly seen
that the
loop contribution is of order $p^2$ and is therefore suppressed by two
powers of the momenta with respect to the tree level result $F^{\pi,tree}_V=1$.
 Notice that $\phi$
 is independent of any
renormalization scale $\mu$ by construction -- the loop integrals
only involve internal as well as external momenta and the masses of the
pions and kaons.
 As long as $d$ is not an integer, the result is finite. When
$d\rightarrow 4$, the loop functions develop a pole,
\beq
\phi(t,M;d)\stackrel{d\rightarrow 4}{\rightarrow}
-\frac{t}{96\pi^2 F^2(d-4)}+O(1) \fs\no
\eeq
 An analogous result holds for all Green functions
evaluated at next--to--leading order: as long as $d$ is different from 4, the
result is finite--ultraviolet singularities only show up when
$d\rightarrow 4$. These singularities are cancelled by the contributions from
the local action at order $p^4$ to which we now turn.

\subsection{Local action at order $p^4$}

To the loop diagrams discussed above, one has to add the tree diagrams from
${\cal L}_2 + {\cal L}_4$ with one vertex from ${\cal L}_4$ and any number of
 vertices from ${\cal L}_2$. The explicit form of ${\cal L}_4$ is \cite{GL1}

\beqa
{\cal L}_4 & = & L_1 \langle D_\mu U^\dagger D^\mu U\rangle^2 +
                 L_2 \langle D_\mu U^\dagger D_\nu U\rangle
                     \langle D^\mu U^\dagger D^\nu U\rangle \no \\*
& & + L_3 \langle D_\mu U^\dagger D^\mu U D_\nu U^\dagger D^\nu U\rangle +
    L_4 \langle D_\mu U^\dagger D^\mu U\rangle \langle \chi^\dagger U +
    \chi U^\dagger\rangle  \no \\*
& & +L_5 \langle D_\mu U^\dagger D^\mu U(\chi^\dagger U + U^\dagger
\chi)\rangle
    +
    L_6 \langle \chi^\dagger U + \chi U^\dagger \rangle^2 +
    L_7 \langle \chi^\dagger U - \chi U^\dagger \rangle^2  \no \\*
& & + L_8 \langle \chi^\dagger U \chi^\dagger U +
 \chi U^\dagger \chi U^\dagger\rangle
    -i L_9 \langle F_R^{\mu\nu} D_\mu U D_\nu U^\dagger +
      F_L^{\mu\nu} D_\mu U^\dagger D_\nu U \rangle \no \\*
& & + L_{10} \langle U^\dagger F_R^{\mu\nu} U F_{L\mu\nu}\rangle +
    L_{11} \langle F_{R\mu\nu} F_R^{\mu\nu} + F_{L\mu\nu} F_L^{\mu\nu}\rangle +
    L_{12} \langle \chi^\dagger \chi \rangle~,
\label{eq:L4}
\eeqa
where
\beqa
  F_R^{\mu\nu} & = & \partial^\mu r^\nu -
                     \partial^\nu r^\mu -
                     i [r^\mu,r^\nu]  \co\\
  F_L^{\mu\nu} & = & \partial^\mu l^\nu -
                     \partial^\nu l^\mu -
                     i [l^\mu,l^\nu] ~. \no
\eeqa
This is the most general Lagrangian of order $p^4$ that exhibits local gauge
invariance
in the form of Eqs.~(\ref{localf}), (\ref{localu}) ,
Lorentz invariance, $P$ and $C$. The
equation of motion has furthermore been used to reduce the number of terms.
 The twelve
new low--energy couplings $L_1, \ldots, L_{12}$ are
divergent (except $L_3$ and $L_7$). They absorb the divergences of the
one--loop
graphs. Consider for illustration again the electromagnetic form factor
$F_V^\pi$. It receives a contribution from $L_9$,
\beq
{F^{\pi,{\cal L}_4}_V (t)}=\frac{2L_9t}{F^2}\fs
\eeq
The sum $F^{\pi,one\;  loop}_V+{F^{\pi,{\cal L}_4}_V}$ of the
one--loop graphs and the counterterm contribution
 is finite, provided $L_9$ is appropriately tuned when
$d\rightarrow 4$,
 \bea
L_9 &=& L_9^r(\mu)+\frac{\lambda(\mu)}{4}\co \label{eq1}\\
\lambda(\mu) &=&(4\pi)^{-2}\mu^{d-4} \left\{ \frac{1}{d-4} - \frac{1}{2}
\left({\mbox{ln}}(4\pi) +\Gamma'(1) +1 \right) \right\} \co\no
\eea
where the renormalized coupling $L_9^r(\mu)$ is finite when $d\rightarrow 4$.
Here we have introduced   a renormalization scale $\mu$.
Although the coupling constant $L_9$ does not know about this scale,
\be\label{eq2}
\mu\frac{dL_9}{d\mu}=0\co
\ee
the
individual pieces in the splitting (\ref{eq1}) do depend on it,
\beq
\mu\frac{d\lambda}{d\mu}=\frac{1}{16\pi^2}+O(d-4)\co\;\;
\mu\frac{dL_9^r}{d\mu}=-\frac{\mu}{4}\frac{d\lambda}{d\mu}\fs
\eeq
Renormalization is now performed by writing
\beq
\phi(t,M;d)=\phi^{ren}(t,M,\mu;d)-\frac{t\lambda(\mu)}{6F^2}\co
\eeq
such that the complete expression for the formfactor
becomes
\be\label{formff}
F_V^\pi(t)=1+2\phi^{ren}(t,M_\pi,\mu;d)+\phi^{ren}(t,M_K,\mu;d)
+\frac{2L_9^rt}{F^2}+O(p^4)\fs
\ee
 $F_V^\pi$ approaches a finite limit when
$d\rightarrow 4$ and is scale independent
by construction. For illustration, we display the renormalized loop integral
\footnote{The combination
$\phi^{ren}(t,M_\pi,\mu;4)+\frac{2L_9^r t}{3F^2}$ is denoted by $H_{\pi\pi}$
in Refs. \cite{semi,glform}.}
at $d=4$,
 \bea\label{eq105}
\phi^{ren}(t,M,\mu;4)&=&
\frac{t}{32\pi^2F^2}\int_0^1\!dx
\;x(1-2x)\log{\left[1-\frac{t}{M^2}x(1-x)\right]}\nl
&-&\frac{t}{192\pi^2F^2}\left\{\log{\frac{M^2}{\mu^2}}+1\right\}\fs
\eea
Note that, as is required by unitarity, the function $\phi$
contains a branch point at $t=4M^2$.
 The expression (\ref{eq105}) is  not expanded further in powers of $t/M^2$,
because this would
result in a Taylor series that is valid only in the range $|t| < 4M^2$.
Stated differently, the momentum and quark expansion is performed at a
fixed value of the ratio $p^2/m_{quark}$.

   The same procedure carries over to all Green functions
collected in the generating functional:
 they remain finite when $d\rightarrow 4$, provided the low--energy
couplings $L_i$ are appropriately tuned in this limit,
\bearr
L_i&=&L_i^r(\mu) +\Gamma_i \lambda(\mu)\fs
\label{eq:div}
\eearr
 The coefficients
$\Gamma_i$ have been evaluated in \cite{GL1}
and are displayed in table \ref{tab:Li}. The couplings $L_i$ are independent of
the
renormalization scale $\mu$ -- therefore, the Green functions do not depend on
it either. The scale dependence of the
renormalized, finite couplings $L_i^r(\mu)$ is governed by the coefficients
$\Gamma_i$,
 \beq
L_i^r(\mu_2) =L_i^r(\mu_1) + \frac{\Gamma_i}{16\pi^2} \ln\frac{\mu_1}{\mu_2}
\; . \label{eq:scale}
\eeq

\begin{table}[t]
\begin{center}
\caption{Phenomenological values and source for the renormalized coupling
constants $L^r_i(M_\rho)$.
The quantities $\Gamma_i$
in the fourth column determine the scale dependence of the $L^r_i(\mu)$
according to Eq.~(\protect\ref{eq:scale}). $L_{11}^r$ and $L_{12}^r$ are not
directly accessible to experiment.} \label{tab:Li}
\vspace{.5cm}
\begin{tabular}{|c||r|l|r|}  \hline
i & $L^r_i(M_\rho) \times 10^3$ & source & $\Gamma_i$ \\ \hline
  1  & 0.4 $\pm$ 0.3 & $K_{e4},\pi\pi\rightarrow\pi\pi$ & 3/32  \\
  2  & 1.35 $\pm$ 0.3 &  $K_{e4},\pi\pi\rightarrow\pi\pi$&  3/16  \\
  3  & $-$3.5 $\pm$ 1.1 &$K_{e4},\pi\pi\rightarrow\pi\pi$&  0     \\
  4  & $-$0.3 $\pm$ 0.5 & Zweig rule &  1/8  \\
  5  & 1.4 $\pm$ 0.5  & $F_K:F_\pi$ & 3/8  \\
  6  & $-$0.2 $\pm$ 0.3 & Zweig rule &  11/144  \\
  7  & $-$0.4 $\pm$ 0.2 &Gell-Mann--Okubo,$L_5,L_8$ & 0             \\
  8  & 0.9 $\pm$ 0.3 & \small{$M_{K^0}-M_{K^+},L_5,$}&
5/48 \\
     &               &   \small{ $(2m_s-m_u-m_d):(m_d-m_u)$}       & \\
 9  & 6.9 $\pm$ 0.7 & $\langle r^2\rangle^\pi_V$ & 1/4  \\
 10  & $-$5.5 $\pm$ 0.7& $\pi \rightarrow e \nu\gamma$  &  $-$ 1/4  \\
\hline
11   &               &                                & $-$1/8 \\
12   &               &                                & 5/24 \\
\hline
\end{tabular}
\end{center}
\end{table}

The constants $F,B$, together with $L_1^r,\ldots,L_{10}^r$,
completely determine the low--energy behaviour of pseudoscalar meson
interactions to $O(p^4)$. $L_{11}^r$ and $L_{12}^r$ are contact terms that
are not directly accessible to experiment. We  discuss the value of
the low--energy constants $L_1^r$,\dots,$L_{10}^r$ below.

\subsection{Chiral anomaly}
The effective Lagrangian ${\cal L}_2 +{\cal L}_4$ is invariant under the local
$SU(3)_L\times SU(3)_R$ transformations (\ref{localf}), (\ref{localu}). Because
dimensional regularization preserves this symmetry, the corresponding
generating functional is invariant as well. On the other hand, the vacuum
transition amplitude is not invariant under the
transformations
(\ref{localf}) -- it is afflicted with anomalies. (In technical terms,
 the relevant fermionic determinant does not allow
for a chiral invariant regularization \cite{FB80}.)
Consider infinitesimal transformations
\beq g_R=1+i\alpha(x)+i\beta(x)\co\; g_L=1+i\alpha(x)-i\beta(x) \fs
\eeq
The conventions in the definition of the fermionic determinant may be chosen
such that the generating functional is invariant under the transformations
generated by the vector currents.
The change in $Z$ then only involves the difference $\beta(x)$ between $g_R$
and $g_L$ \cite{bardeen,WZ},
 \bea\label{anomaly}
\delta Z &=& -\int dx \langle \beta(x)\Omega(x)\rangle\co\nl
\Omega(x)&=&\frac{N_c}{16\pi^2}\epsilon^{\alpha\beta\mu\nu}\left
[v_{\alpha\beta} v_{\mu\nu} +\frac{4}{3}D_\alpha a_\beta D_\mu a_\nu
+\frac{2i}{3}\{v_{\alpha\beta},a_{\mu a_\nu}\} \right.\nl
&&\hspace{1.9cm}+\left.\frac{8i}{3}a_\mu
v_{\alpha\beta}a_\nu   +\frac{4}{3}a_\alpha a_\beta a_\mu a_\nu\right ]\co\nl
v_{\alpha\beta}&=&\partial_\alpha v_\beta -\partial_\beta v_\alpha
-i[v_\alpha,v_\beta]\co\nl
D_\alpha a_\beta &=& \partial_\alpha a_\beta -i[v_\alpha,a_\beta]\fs
\eea
($N_c$ is the number of colours and  $ \varepsilon_{0123} = 1$.)
Notice that $\Omega$ only depends on the external
fields $v_\mu$ and $a_\mu$. The quark masses do not occur -- the anomalies are
independent thereof.
A functional $Z[U,l,r]$ that reproduces the chiral anomaly
was first constructed by Wess and Zumino \cite{WZ}. For practical
purposes, it is useful to write it in the explicit form given by Witten
 \cite{Witten}:
\beqa
Z[U,l,r]_{WZW} &=&-\dfrac{i N_c}{240 \pi^2}
\int_{M^5} d^5x \epsilon^{ijklm} \langle \Sigma^L_i
\Sigma^L_j \Sigma^L_k \Sigma^L_l \Sigma^L_m \rangle \label{eq:WZW} \\*
 & & - \dfrac{i N_c}{48 \pi^2} \int d^4 x
\varepsilon_{\mu \nu \alpha \beta}\left( W (U,l,r)^{\mu \nu
\alpha \beta} - W ({\bf 1},l,r)^{\mu \nu \alpha \beta} \right)
\no \\
W (U,l,r)_{\mu \nu \alpha \beta} & = &
\langle U l_{\mu} l_{\nu} l_{\alpha}U^{\dg} r_{\beta}
+ \frac{1}{4} U l_{\mu} U^{\dg} r_{\nu} U l_\alpha U^{\dg} r_{\beta}
+ i U \partial_{\mu} l_{\nu} l_{\alpha} U^{\dg} r_{\beta}
\no  \\
& & + i \partial_{\mu} r_{\nu} U l_{\alpha} U^{\dg} r_{\beta}
- i \Sigma^L_{\mu} l_{\nu} U^{\dg} r_{\alpha} U l_{\beta}
+ \Sigma^L_{\mu} U^{\dg} \partial_{\nu} r_{\alpha} U l_\beta
\no \\
& & -\Sigma^L_{\mu} \Sigma^L_{\nu} U^{\dg} r_{\alpha} U l_{\beta}
+ \Sigma^L_{\mu} l_{\nu} \partial_{\alpha} l_{\beta}
+ \Sigma^L_{\mu} \partial_{\nu} l_{\alpha} l_{\beta}  \\
& & - i \Sigma^L_{\mu} l_{\nu} l_{\alpha} l_{\beta}
+ \frac{1}{2} \Sigma^L_{\mu} l_{\nu} \Sigma^L_{\alpha} l_{\beta}
- i \Sigma^L_{\mu} \Sigma^L_{\nu} \Sigma^L_{\alpha} l_{\beta}\rangle
\no \\
& & - \left( L \leftrightarrow R \right) \co\no \eeqa
$$
\Sigma^L_\mu = U^{\dg} \partial_\mu U \co\;\;
\Sigma^R_\mu = U \partial_\mu U^{\dg} $$
$$ N_c = 3 \co $$
where $\left( L \leftrightarrow R \right)$ stands for the interchange
$$
U \leftrightarrow U^\dg~, \qquad l_\mu \leftrightarrow r_\mu~~,
\qquad \Sigma^L_\mu \leftrightarrow \Sigma^R_\mu ~. $$
The first term in Eq.~(\ref{eq:WZW}) bears the mark of the
anomaly: this part of the action is local in {\it five} dimensions,
but it cannot be
written as a finite polynomial in $U$ and $\partial_\mu U$ in four dimensions.
This term involves at least five pseudoscalar
fields and will not be needed in the following chapters. It does
contribute to $K_{e5}$ decays, but its contribution
is proportional to the electron mass and therefore strongly suppressed,
see the section on $K_{e5}$ decays
 \cite{semi}. The convention used in Eq.~(\ref{eq:WZW}) ensures
that $Z[U,l,r]_{WZW}$ conserves parity and reproduces the transformation law
(\ref{anomaly}).

The Wess--Zumino--Witten functional contains all the
anomalous contributions to electromagnetic and semileptonic weak meson
decays. The relevant piece for, e.g., $K_{l4}$ decays is
\beq
Z[U,l,r]_{WZW}=\frac{i\sqrt{2}}{ 4 \pi^2 F^3}
\int d^4 x \epsilon_{\mu \nu \rho \sigma}\langle\partial^\mu \Phi \partial^\nu
\Phi
\partial^\rho \Phi v^\sigma\rangle + \cdots \; \; .
\eeq

Furthermore, all Ward identities originating from the
$SU(3)_L\times SU(3)_R$  transformation properties of the underlying theory
may be derived from
the fact  that i) the difference $Z-Z_{WZW}$ is invariant under  local gauge
transformations, and ii)  $Z_{WZW}$
 transforms in the manner indicated in Eq.~(\ref{anomaly}).

\subsection{The low--energy constants $L_i$}
\label{sec3}
Similarly to $F$ and $B$ in the lowest--order Lagrangian ${\cal L}_2$,
the low--energy constants $L_i^r$ are not constrained
by chiral symmetry -- they are fixed by the dynamics of the underlying theory
in terms of the renormalization group invariant scale $\Lambda$ and the heavy
quark masses $m_c,m_b,\ldots \; .$ With present techniques, it is however not
possible to evaluate them directly from the QCD Lagrangian (for some
attempts in this direction see Ref.~\cite{models}). Therefore, they have
been determined by comparison with
experimental low--energy information and by using large--$N_c$ arguments.
The result is shown in column 2 of table \ref{tab:Li}, where
$L_1^r,\ldots,L_{10}^r$ are
displayed at the scale $\mu=M_{\rho}$. The experimental information
underlying these values is shown in column 3.
$L_1,L_2$ and $L_3$ are taken from a recent overall
fit to $K_{e4}$ and $\pi \pi$ data \cite{BCG}, see also Refs. \cite
{Kl4Bij,Kl4Rig} and the section on $K_{l4}$ decays in \cite{semi}.
$L_4,\ldots,L_{10}$ are from \cite{GL1}. For $L_9$ see also \cite{BC88}.
The combination $L_9 + L_{10}$ has also been determined in
\cite{Bab92} from data on $\gamma\gamma \to \pi^+\pi^-$.

Once the low--energy couplings $L_i^r$ are known, the machinery set up above
allows one to make predictions. As a simple example consider the charge radius
of the electromagnetic form factor,
\bea
F_V^\pi(t)&=&1+\frac{1}{6}\langle r^2\rangle^\pi_V t + O(t^2) \fs
\eea
The expression for $\langle r^2\rangle^\pi_V$ follows easily from
Eqs.~(\ref{formff}) and (\ref{eq105}),
\bea\label{radius}
\langle r^2\rangle^\pi_V=
\frac{12 L_9^r}{F^2}-\frac{1}{32\pi^2F^2}\left\{2\log{\frac{M_\pi^2}{\mu^2}}+
\log{\frac{M_K^2}{\mu^2}} +3\right\} +O(m_{quark})\fs
\eea
Here, the symbol $O(m_{quark})$ stands for higher--order contributions which
are not worked out here (two loops and beyond).
Therefore, the constant $L_9^r$ may be pinned down from the
experimental value \cite{r2em} $\langle r^2\rangle^\pi_V=0.439\pm 0.008$
fm$^2$. The same constant occurs \cite{semi,glform} in
the slope $\lambda_+$ of the form factor $f_+^{K\pi}(t)$ measured in $K^0_{e3}$
decay \cite{pdg94},
\beq
\lambda_+=0.0286\pm 0.0022\fs
\eeq
Inserting  $L_9^r=6.9\cdot 10^{-3}$ in the expression for $\lambda_+$
 gives for the central value  the postdiction
\beq
\lambda_+=0.031\fs
\eeq
We see that the machine works properly. For a more thorough discussion of
charge radii in the meson sector we refer the reader to Refs.
\cite{semi,glform}.  Many more predictions
have been worked out -- for reviews see, e.g., Ref.~\cite{reviews}.
The semileptonic decays which will be measured with high precision at
DA$\Phi$NE  provide the opportunity to expose the method to very
thorough consistency checks. We display for this purpose in table
\ref{tab:semili}
the low--energy couplings that occur in the matrix elements of the semileptonic
kaon decays discussed in \cite{semi}.
(There is an ambiguity concerning the bookkeeping of $L_4$ and $L_5$: some of
these contributions may be absorbed into the physical decay constants
$F_\pi,F_K$. Here we have chosen the convention that corresponds to the
amplitudes displayed in \cite{semi}.
Furthermore, in $K_{e5}$ decays, additional constants  may occur.
The relevant one--loop corrections have however not been worked
out yet.
This channel is therefore omitted in the table.)
\begin{table}[t]
\protect
\begin{center}
\caption{Occurrence
of the low--energy coupling constants $L_1,\ldots,L_{10}$ and of
the anomaly  in the semileptonic decays discussed
in {\protect\cite{semi}}. In $K_{\mu 4}$ decays,
the same constants as in the electron mode (displayed here) occur. In
addition, $L_6$ and $L_8$ enter in the channels
$K^+\rightarrow \pi^+\pi^-\mu^+\nu_\mu$ and
$K^+\rightarrow \pi^0\pi^0\mu^+\nu_\mu.$ \label{tab:semili}   } \vspace{1em}
\begin{tabular}{|c||c|c|c|c|c|c|c|}  \hline
 & & & & & $K^+\rightarrow$ & $K^+\rightarrow$ & $K^0\rightarrow$
\\
     &
$K_{l2\gamma}$ &
$K_{l2ll}$    &
$K_{l3}$    &
$K_{l3\gamma}$&
$\pi^+\pi^-e^+\nu_e$&
$\pi^0\pi^0e^+\nu_e$&
$\pi^0\pi^-e^+\nu_e$ \\ \hline
% 1     times2 times3 times4 times5 times6  times7  times8
 $L_1$ &      &      &      &      & $\times$ &$\times$ &
\\
 $L_2$&      &      &      &      & $\times$ &$\times$ &
\\
 $L_3$ &      &      &      &      & $\times$ &$\times$ & $\times$
\\
 $L_4$ &      &      &      &      & $\times$ &$\times$ &
\\
 $L_5$&       &      &      &      & $\times$ &$\times$ &$\times$
\\
 $L_9$&      &$\times$&$\times$&$\times$& $\times$ &$\times$ &$\times$
\\
 $L_9+L_{10}$ &$\times$&$\times$&      &$\times$&       & &
\\
\hline
Anomaly&$\times$&$\times$& &$\times$&$\times$&&$\times$
\\
\hline
\end{tabular}
\end{center}
\end{table}

Are these values of the low--energy constants
consistent with the general picture
 underlying the chiral perturbative analysis or are they unduly large?
Consider the charge radius of the electromagnetic form factor worked out above.
 If the pion and kaon cuts were the only low--energy singularities of
importance, we would expect the main contribution to the charge radius to stem
from the chiral logarithms and the contribution from $L_9$ to be negligible.
With a scale $\mu$ somewhere in the range from
500 MeV to 1 GeV gives $\langle r^2\rangle^\pi_V =0.03\ldots 0.09$ fm$^2$
from the curly
bracket in Eq.~(\ref{radius}), clearly outside the experimental value.
 Therefore, the coupling $L_9$ generates
the main contribution. On the other hand,
the observed value of the radius is consistent with $\rho$ dominance,
$\langle r^2\rangle^\pi_V \simeq 6M_\rho^{-2}=0.4$ fm$^2$. To understand the
size of $L_9$ one therefore needs to understand how the presence of an excited
$q\bar q$ state at $M_\rho=$770 MeV affects the low--energy structure of the
Green functions \cite{GL1}. This will be discussed in some detail in
subsection 3.

\subsection{Chiral  power counting}
\label{sec:power}
 Power counting in the low--energy expansion may be discussed in a concise
manner as follows \cite{Wein79}.
 Take a fixed term in the expansion
 of the generating functional in powers of the external
fields -- an example is given in Eq.~(\ref{zpipi}).
Let us further consider a fixed, connected $L$--loop diagram that contributes
to this term.   Since all internal
lines  correspond to pseudoscalar mesons, the finite part of the
loop integral is a homogeneous
function of the external momenta, the meson masses and the renormalization
scale $\mu$.
Denote by $D_L$  the degree of homogeneity, and by $D_F$ the
number of external fields times their dimension according
 to (\ref{counting}) (i.e., $D_F=4$ in the present case). We call
\beq
D=D_L+D_F
\eeq
the {\it chiral dimension} of the $L$--loop amplitude.

  For a general connected $L$--loop diagram with
$N_d$ vertices of $O(p^d)$ ($d = 2,4,\ldots)$, it is given by \cite{Wein79}
\beq
D = 2L + 2 + \sum_d (d-2) N_d~, \qquad d = 2,4,\ldots \label{DL}
\eeq
and therefore, up to $O(p^4)$ :
\beqa
D = 2 : & L = 0, \; d = 2 \qquad & Z_2 = \int d^4x {\cal L}_2 \nl
D = 4 : & L = 0, \; d = 4 \qquad & Z_4^{\rm tree} = \int d^4x {\cal
L}_4+Z_{WZW}\nl & L =1, \; d = 2 \qquad & Z_4^{L=1} \mbox{ for } {\cal L}_2~.
\eeqa
To verify, we consider the case of the $\pi\pi$ scattering amplitude at tree
level: in addition to the amplitude (\ref{apipi}) of homogeneity 2, the
 Green function (\ref{zpipi}) contains four external propagators
and four (derivative) couplings to the axial currents. Therefore,
$D_L=2+4*(-2+1)=-2$, $D_F=4$, such that $D=2$.

This is true for each term in the expansion (\ref{zpipi}), such that
\be
 Z=Z_2+Z_4^{\rm tree}+Z_4^{L=1} + O(p^6)\fs
\ee

For a given Green function, the chiral dimension $D$ increases with
$L$ according to Eq.~(\ref{DL}). In order to reproduce the (fixed)
physical dimension of the amplitude, each loop produces a factor $1/F^2$.
Together with the geometric loop factor $(4\pi)^{-2}$, the loop expansion
suggests
\beq
4\pi F_\pi = 1.2 \mbox{ GeV}
\eeq
as the natural scale of the chiral expansion \cite{GeMa}. Restricting
the domain of applicability of CHPT to momenta $|p|\; \lets \;
O(M_K)$, the natural expansion parameter of chiral amplitudes is
expected to be of the order
\beq
\frac{M_K^2}{16 \pi^2 F_\pi^2} = 0.18~.
\eeq
In addition, these terms often occur multiplied with chiral logarithms. It is,
therefore, no surprise that substantial higher--order
 corrections  in the chiral expansion are the rule rather than the exception
in the framework of $SU(3)\times SU(3)$.
On the other hand, for $SU(2)\times SU(2)$ and for momenta
$|p|\; \lets \;O(M_\pi)$ the chiral expansion is expected to converge
considerably faster.

%before each section
\setcounter{equation}{0}
\setcounter{subsection}{0}

\section{Meson resonances in CHPT}

We now discuss the singularities generated by the exchange of nearby
resonances like the rho, omega, phi etc. and their effect on the low--energy
structure of the Green functions.
A systematic analysis of the couplings between meson resonances of the
type $V$, $A$, $S$, $P$ and the pseudoscalar mesons was performed
in Ref. \cite{EGPR} (see
also Ref. \cite{Resref} for related work).

It is straightforward to couple general matter fields to the
Goldstone modes using the methods of non--linear realizations of
the chiral group $G = SU(3)_L\times SU(3)_R$ \cite{CCWZ}. The chiral
transformation properties of the resonance fields
depend only on their transformation
properties under the diagonal subgroup $SU(3)_V$. A non--linear
realization of spontaneously broken chiral symmetry is defined
\cite{CCWZ} by specifying the action of $G$ on
the elements $u(\phi)$ of the coset space $SU(3)_L\times SU(3)_R/SU(3)_V$:
\beq
u(\phi) \stackrel{G}{\ra} g_R\, u(\phi)\,
h(g,\phi)^\dagger \, = \, h(g,\phi)\, u(\phi)\, g_L^\dagger \, ,
\label{eq:nlr}\eeq
where $\phi^i\, (1\leq i\leq 8)$ are the Goldstone fields of
Eq.~(\ref{eq:phi}). In the usual convention, the relation between
$u(\phi)$ and $U(\phi)$ in Eq.~(\ref{eq:phi}) is simply
\beq
U(\phi) \, = \, u(\phi)^2 \, .
\ee

The so--called compensating $SU(3)_V$ transformation $h(g,\phi)$ defined by
Eq.~(\ref{eq:nlr}) allows the construction of an arbitrary non--linear
realization of $G$. For instance, an octet
\beq
R = \frac{1}{\sqrt{2}} \, \lambda_a R^a
\eeq
of resonance fields transforms as
\beq
R \stackrel{G}{\ra} h(g,\phi) R h(g,\phi)^\dagger ~.\label{eq:hom}
\eeq
Here, we shall only discuss $V$ and $A$ resonances, referring to \cite{EGPR}
for all details including the treatment of $S$ and $P$ resonances.

\subsection{Antisymmetric tensor field formulation}

 We describe the $V$ and $A$ mesons  by antisymmetric tensor fields
$R_{\mu\nu}=-R_{\nu\mu}$ \cite{GL1,EGPR}.
Alternative treatments of spin--1 fields are considered below.
The tensor field formulation is especially convenient for including
the interactions with external gauge fields.
Another advantage is that even in the presence of interactions the spin--1
character of the field is not modified. This is in contrast to the
vector field formulation where couplings of the form
$V_\mu \partial^\mu S$ with a scalar field $S$ may arise requiring a
redefinition of the spin--1 vector field. A well--known example
of this type is $a_1$--$\pi$ mixing.

 It turns
out \cite{EGPR} that for $V$ and $A$ resonances only octets can
contribute to the $L_i$.
For the kinetic terms we define a covariant derivative
\beq
\nabla_\mu R \,=\, \partial_\mu R + [\Gamma_\mu,R] \, ,
\eeq
with a connection
\beq
\Gamma_\mu \,=\,
\frac{1}{2} \{ u^\dagger (\partial_\mu - i r_\mu )u +
  u(\partial_\mu - i l_\mu ) u^\dagger \} \label{eq:conn}
\eeq
ensuring the proper transformation
\beq
\nabla_\mu R \stackrel{G}{\ra} h(g,\phi)\,
\nabla_\mu R \, h(g,\phi)^\dagger\, .
\eeq
Invoking $P$ and $C$ invariance, one finds that all couplings linear
in the fields $V$, $A$, $S$ and $P$ start at order $p^2$. The complete
resonance Lagrangian of lowest order \cite{EGPR}
\beq
{\cal L}_\mscr{res} \, =\,
\sum_{R=V,A,S,P} \{  {\cal L}_\mscr{kin}(R) + {\cal L}_2(R)\} \, ,
\label{eq:Lres} \eeq
consists of kinetic terms
\beq
{\cal L}_\mscr{kin}(R) =
    - \dfrac{1}{2}\, \langle \nabla^\lambda R_{\lambda\mu}
\nabla_\nu R^{\nu\mu} - \dfrac{M^2_R}{2} \, R_{\mu\nu} R^{\mu\nu} \rangle
\qquad\quad R = V,A~,
\label{eq:Rkin}
\eeq
where $M_R$ are the corresponding masses in the chiral limit,
and of interaction Lagrangians
\beqa
{\cal L}_2[V(1^{--})] &=& \dfrac{F_V}{2 \sqrt{2}} \,
     \langle V_{\mu\nu} f_+^{\mu\nu}\rangle +
    \dfrac{iG_V}{\sqrt{2}} \, \langle V_{\mu\nu} u^\mu u^\nu\rangle
\label{eq:RintV}  \\
{\cal L}_2[A(1^{++})] &=& \dfrac{F_A}{2 \sqrt{2}} \,
    \langle A_{\mu\nu} f_-^{\mu\nu} \rangle \label{eq:RintA} \\
f^{\mu\nu}_\pm & = & u F_L^{\mu\nu} u^\dagger \pm u^\dagger
F_R^{\mu\nu} u ~. \no
\eeqa
The coupling constants $F_V$, $G_V$ and $F_A$ are real. As
already mentioned, we only exhibit the $V$ and $A$ Lagrangians here.

The coupling constants $F_V$, $G_V$ and $F_A$ (and the corresponding ones for
$S$, $P$ resonances) can be estimated from resonance decays,
see \cite{EGPR} for details. We use
\beqa
|F_V|=154 \mbox{ MeV} & \qquad & |G_V|= 53 \mbox{ MeV} \no \\
|F_A|= 123 \mbox{ MeV}& \qquad & M_A= 968 \mbox{ MeV}
\eeqa
together with $M_V = M_{\rho}$.

Resonance exchange (more specifically, here $V$ and $A$ exchange only)
then gives rise to the following contributions to the $L_i$ \cite{EGPR}~:
\beq
\ba{lll}
L_1^V = \dfrac{G_V^2}{8 M_V^2} \qquad & L_2^V = 2 L_1^V \qquad
& L_3^V = - 6 L_1^V \\[10pt]
L_9^V = \dfrac{F_V G_V}{2 M_V^2} \qquad & L_{10}^V = - \dfrac{F_V^2}{4 M_V^2}
\qquad & L_{10}^A = \dfrac{F_A^2}{4 M_A^2} ~.
\ea \label{eq:VAex}
\eeq

The  resulting $L_i$ are summarized in table~\ref{tab:vmd},
which compares the different re\-son\-ance ex\-change contributions with the
phenomenologically determined values of the $L_i^r(M_\rho)$. For vector and
axial--vector mesons only the $SU(3)$ octets contribute, whereas both octets
and singlets are relevant in the case of scalar and pseudoscalar resonances.

%%%%%%%%%%%%%%%%%%%%%%%%%%% TABLE %%%%%%%%%%%%%%%%%
\begin{table}
\caption{$V$, $A$, $S$, $S_1$ and $\eta_1$ contributions to the
 coupling constants $L_i^r$ in units of $10^{-3}$.
 The last column shows the results obtained with the relations
 (\protect\ref{eq:Lminmod}) in the minimal model of $V, A$ resonance
 couplings.}
\vspace{0.2cm}
\label{tab:vmd}
\begin{center}
\begin{tabular}{|c|r||cllllc|r||r|}
\hline i & $L_i^r(M_\rho)$ && $V$ & $A\,$ & $\,S$ &
      $S_1$ & $\eta_1$ & Total & MM$^{c)}$
\\ \hline 1 & $0.4\pm0.3$ &\hspace{3mm} &
      $0.6$ & $0$ & $\hspace{-3.3mm}-0.2$ &
      $0.2^{b)}$ & $0$ & $0.6$ & $0.9$
\\ 2 & $1.35\pm0.3$ &\hspace{3mm} &
      $1.2$ & $0$ & $0$ &
      $0$ & $0$ & $1.2$ & $1.8$
\\ 3 & $-3.5\pm1.1$ &\hspace{3mm}  &
      $\hspace{-3.3mm}-3.6$ & $0$ & $0.6$ &
      $0$ & $0$ & $-3.0$ & $-4.9$
\\ 4 & $-0.3\pm0.5$ &\hspace{3mm}  &
      $0$ & $0$ & $\hspace{-3.3mm}-0.5$ &
      $0.5^{b)}$ & $0$ & $0.0$ & $0.0$
\\ 5 & $1.4\pm0.5$ & \hspace{3mm}  &
      $0$ & $0$ & $1.4^{a)}$ &
      $0$ & $0$ & $1.4$ & $1.4$
\\ 6 & $-0.2\pm0.3$ &\hspace{3mm}   &
      $0$ & $0$ & $\hspace{-3.3mm}-0.3$ &
      $0.3^{b)}$ & $0$ & $0.0$ & $0.0$
\\ 7 & $-0.4\pm0.2$ &\hspace{3mm}    &
      $0$ & $0$ & $0$ &
      $0$ & $-0.3$ & $-0.3$ & $-0.3$
\\ 8 & $0.9\pm0.3$ & \hspace{3mm}     &
      $0$ & $0$ & $0.9^{a)}$ &
      $0$ & $0$ & $0.9$ & $0.9$
\\ 9 & $6.9\pm0.7$ & \hspace{3mm}      &
      $6.9^{a)}$ & $0$ & $0$ &
      $0$ & $0$ & $6.9$ & $7.3$
\\ 10 & $-5.5\pm0.7$ &\hspace{3mm}      &
      $\hspace{-5.3mm}-10.0$ & $4.0$ & $0$ &
      $0$ & $0$ & $-6.0$ & $-5.5$
\\ \hline
\end{tabular}

\hbox{$\qquad\qquad\qquad\quad$ $^{a)}$ Input. $\qquad$
$^{b)}$ Large--$N_c$ estimate. $\qquad$
$^{c)}$ With (\protect\ref{eq:Lminmod})}
\end{center}
\end{table}
%%%%%%%%%%%%%%%%%%%%%%%%%%%%%%%%%%%%%%%%%%%%%%%%%%%%%%%%%%%%%%%%

The results displayed in table \ref{tab:vmd} can be summarized
in the following way:
\begin{description}
\item[Chiral duality:] \mbox{ } \\
The $L_i^r(M_\rho)$ are practically saturated by resonance exchange.
There is very little room left for additional contributions.
\item[Chiral VMD:] \mbox{ } \\
Whenever spin--1 resonances can contribute at all ($i = 1,2,3,9,10$),
the $L_i^r(M_\rho)$ are almost completely dominated by $V$ (and for
$L_{10}$ only, also $A$) exchange. In particular, the strong $V,A$ exchange
contributions are responsible for the relatively large coupling constants
$L_3$, $L_9$ and $L_{10}$.
\end{description}
\noindent

\subsection{Alternative descriptions for spin--1 fields}

The use of antisymmetric tensor fields to describe spin--1 resonances
is just a matter of convenience. In fact, the high--energy structure
of QCD allows to establish a complete equivalence between different
formulations and it gives rise to additional information \cite{EGLPR} ~:
\bit
\item Imposing the QCD constraints at high energies via
dispersion relations,
all phenomenologically successful models for $V,A$ resonances are
equivalent to $O(p^4)$: tensor field description used in
Eqs.~(\ref{eq:Rkin}), (\ref{eq:RintV}), (\ref{eq:RintA}), massive Yang--Mills
\cite{Mei}, hidden--gauge formulations \cite{Bando}, etc.
\item With additional QCD--inspired assumptions of high--energy behaviour,
like an unsubtracted dispersion relation for the pion form factor, all
$V$ and $A$ couplings can be expressed in terms of $F_\pi$ and $M_\rho$
only:
\beqa
F_V=\sqrt{2} F_\pi &\qquad&
G_V= F_\pi /\sqrt{2} \no\\*
F_A = F_\pi &\qquad& M_A = \sqrt{2} M_V~.
\label{eq:minmod}
\eeqa
The spin--1 resonance exchange contributions to the $O(p^4)$ chiral couplings
are in this case
\beq
L_1^V = L_2^V/2 = - L_3^V/6 = L_9^V/8 = -L_{10}^{V+A}/6 = F_\pi^2/(16 M_V^2)~.
\label{eq:Lminmod}  \\[5pt]
\eeq
The last column in table~\ref{tab:vmd} shows the predictions for
the $L_i$ using the relations (\ref{eq:Lminmod}). The agreement
is quite remarkable.
\eit

\noindent
The theoretical prediction $G_V = F_\pi/\sqrt{2}$ implies
\beq
\Gamma(\rho \ra 2\pi) = \frac{M_\rho^3(1 - 4M_\pi^2/M_\rho^2)^{3/2}}
{96 \pi F_\pi^2} = 141 \mbox{ MeV }~,
\eeq
which is one version of the so--called KSFR relation \cite{KSFR}. This
is a non--trivial result: higher--order terms in the chiral expansion
could a priori influence the decay $\rho \ra 2\pi$, but would
be negligible
at small momenta $|p| \;\lets\; M_K$. The underlying reason for the
success of the KSFR relation is once again the intimate connection between
the high--energy structure predicted by QCD and the low--energy structure
of CHPT via dispersion relations \cite{Eckrho}.

The most compact way to exhibit the relations (\ref{eq:minmod}) in
a Lagrangian form is by way of a ``minimal model''
\cite{EGLPR} employing vector fields for the spin--1 resonances
\cite{Mei,Bando} :
\beqa
{\cal L}_{MM} & = & -{1\over 4}\langle \bar V_{\mu\nu} \bar V^{\mu\nu}\rangle
+ {1\over 2}
M_V^2 \,\langle \left(\bar V_\mu -{2 i F_\pi\over M_V}\Gamma_\mu\right)^2
\rangle  \label{eq:LMM} \\
& & -{1\over 4}\langle \hat A_{\mu\nu} \hat A^{\mu\nu}\rangle
+ M_V^2 \,\langle \left(\hat A_\mu +{F_\pi\over 2 M_V} u_\mu\right)^2
\rangle \no \\
\bar V_{\mu\nu} & = & \partial_\mu\bar V_\nu - \partial_\nu\bar V_\mu -
i {M_V \over 2 F_\pi} \,[\bar V_\mu,\bar V_\nu ] \no\\
\hat A_{\mu\nu} & = & \nabla_\mu\hat A_\nu - \nabla_\nu\hat A_\mu \no~.
\eeqa
The vector field $\bar V_\mu$ transforms like a gauge field
under chiral rotations,
\beq
\bar V_\mu \stackrel{G}{\ra} h(g,\phi) \,\bar V_\mu\, h(g,\phi)^\dagger
+ {2 i F_\pi\over M_V}\, h(g,\phi)\,\partial_\mu h(g,\phi)^\dagger ~,
\label{eq:Vhat} \eeq
whereas the axial--vector field $\hat A_\mu$ transforms homogeneously
as in Eq.~(\ref{eq:hom}).

\vspace*{2cm}

This short introduction to CHPT (see Refs. \cite{reviews,books} for more
extensive treatments with references to the original literature)
contains all the ingredients necessary
for the calculation of semileptonic $K$ decay amplitudes to $O(p^4)$
\cite{semi}.
The chiral realization of the
non--leptonic weak interactions is discussed in \cite{DEIN}.

\newpage
%The following line only refers to the preprint version
%\addcontentsline{toc}{section}{\hspace{1cm}Bibliography}

\end{document}